\documentclass[a4paper]{article}

\usepackage{INTERSPEECH2020}
\usepackage{txfonts}
\usepackage{here}
\usepackage{url}
\usepackage{cite}
\usepackage{colortbl}
\usepackage{setspace}

\usepackage{multirow}
\usepackage{tabularx}
\newcolumntype{C}[1]{>{\centering\arraybackslash}p{#1}}
\newcolumntype{L}[1]{>{\raggedright\arraybackslash}p{#1}}
\newcolumntype{R}[1]{>{\raggedleft\arraybackslash}p{#1}}

\title{Exploring the Best Loss Function for DNN-Based Low-latency Speech Enhancement with Temporal Convolutional Networks}
\name{Yuichiro Koyama$^{12}$, Tyler Vuong$^1$, Stefan Uhlich$^3$ and Bhiksha Raj$^1$}
\address{
  $^1$Carnegie Mellon University, Pittsburgh, PA, USA\\
  $^2$Sony Corporation, Tokyo, Japan \hspace{1cm}
  $^3$Sony Europe B.V., Stuttgart, Germany}
\email{Yuichiro.Koyama@sony.com, tvuong@andrew.cmu.edu, Stefan.Uhlich@sony.com, bhiksha@cs.cmu.edu}

\begin{document}
\setlength{\abovedisplayskip}{5pt} 
\setlength{\belowdisplayskip}{5pt}

\maketitle
\begin{spacing}{0.9}
\begin{abstract}
Recently, deep neural networks (DNNs) have been successfully used for speech enhancement, and DNN-based speech enhancement is becoming an attractive research area.
While time-frequency masking based on the short-time Fourier transform (STFT) has been widely used for DNN-based speech enhancement over the last years, time domain methods such as the time-domain audio separation network (TasNet) have also been proposed.
The most suitable method depends on the scale of the dataset and the type of task.  
In this paper, we explore the best speech enhancement algorithm on two different datasets.
We propose a STFT-based method and a loss function using problem-agnostic speech encoder (PASE) features to improve subjective quality for the smaller dataset.
Our proposed methods are effective on the Voice Bank + DEMAND dataset and compare favorably to other state-of-the-art methods. 
We also implement a low-latency version of TasNet, which we submitted to the DNS Challenge and made public by open-sourcing it.
Our model achieves excellent performance on the DNS Challenge dataset.   
\end{abstract}
\noindent\textbf{Index Terms}: deep learning, speech enhancement, perceptual quality

\section{Introduction}
Speech enhancement, which removes noise signals from the observed signal in order to enhance the speech signal, is one of the most useful technologies not only  to improve the performance of various speech related applications such as speech recognition but also to enable people to listen and understand recorded speech more clearly. Recently, deep neural networks (DNNs) have been successfully used in speech enhancement, and DNN-based speech enhancement is becoming an attractive research area.
Due to these recent developments, the Deep Noise Suppression (DNS) Challenge~\cite{reddy2020interspeech} was organized together with this year's Interspeech. 
This competition compared single-channel  speech enhancement systems with respect to their (perceptual) quality of the enhanced speech.

Over the last years, time-frequency masking based on the short-time Fourier transform (STFT) has been widely used for DNN-based speech enhancement~\cite{narayanan2013ideal,wang2014structure,wang2014on,soni2018time,xia2020weighted}. 
While approaches that estimate not only the amplitude but also the phase have been proposed~\cite{takahashi2018phasenet,williamson2015complex}, in~\cite{wisdom2019differentiable}, a framework that optimizes the network such that the separated signals satisfy both STFT consistency and mixture consistency was proposed for real-valued as well as complex-valued masks. 
Also, the Phase-and-Harmonics-Aware Speech Enhancement Network (PHASEN)~\cite{yin2019phasen} has achieved high performance using a two-stream architecture, which efficiently estimates the amplitude and phase while the streams exchange information with each other.
All these approaches are based on processing the signals in the STFT domain.

On the other hand, time domain methods have also been studied~\cite{pascual2017segan,rethage2018wavenet,germain2018speech}. 
In particular, the Time-domain audio separation network (TasNet)~\cite{luo2018tasnet} and its fully convolutional version (Conv-TasNet)~\cite{luo2019conv}, which have a trainable encoder-decoder architecture and do not depend on the STFT, have  surpassed the performance of the ideal binary mask (IBM) for speech separation; in other words, it has exceeded the upper limit of the performance of time-frequency masking. 
Moreover, architectures using a temporal convolutional network (TCN)~\cite{lea2016temporal,lea2017temporal,bai2018empirical} such as Conv-TasNet are considered to be more suitable than recurrent neural network architectures for real-time implementation as they are unaffected by the start point
and can be operated efficiently~\cite{luo2019conv, bai2018empirical}.

The advantage of using the STFT is that the knowledge obtained by  time-frequency analysis in past research can be applied to the architecture. For example, PHASEN takes into account the harmonic characteristic of speech~\cite{yin2019phasen}. 
If enough data is available, on the other hand, the performance of time domain methods such as Conv-TasNet is expected to be higher than that of time-frequency masking because a projection to an appropriate domain can be learned from the distribution of the training data. Therefore, it is necessary to determine which method is the best by considering the scale of the dataset.

In this paper, we explore the best speech enhancement algorithm and the best loss function on both the Voice Bank + DEMAND (VBD) dataset~\cite{veaux2013voice,thiemann2013diverse,valentini2016investigating}, which is widely used to evaluate speech enhancement algorithms, and the dataset provided for the DNS Challenge (DNS dataset)~\cite{reddy2020interspeech}. 
We also compare several types of loss functions including the use of the problem-agnostic speech encoder (PASE)~\cite{Pascual2019,ravanelli2020multi}  to improve objective measures highly related to the subjective quality~\cite{hu2007evaluation}.
The contribution of this paper is threefold. 
First, we propose a novel method called STFT-TCN, which is based on Conv-TasNet and utilizes the STFT/ISTFT for the encoder/decoder instead of trainable parameters, similarly to \cite{heitkaemper2019demystifying}. 
Since the size of the VBD dataset is relatively small, we can expect that fixing the encoder and decoder can simplify the problem that the DNN has to solve and achieve better performance. 
It achieved comparable performance to other state-of-the-art methods on the VBD dataset in terms of the metrics highly correlated with the subjective quality. 
Second, we found that using PASE in the loss function improves the metrics correlated with subjective quality on the VBD dataset.  Previous works~\cite{Pascual2019,ravanelli2020multi,Alvarez2019} use PASE as a feature extractor for downstream tasks such as speech recognition, however we only use PASE during training when computing the loss to optimize our model.
Third, we extend our best scheme that was found on the VBD dataset for the DNS Challenge and our modified Conv-TasNet, which we submitted to the DNS Challenge, achieves the best performance.

\vspace{-3mm}
\end{spacing}
\begin{spacing}{0.9}
\section{Related works}

The discrete-time signal captured by a microphone can be written as $x(n) = s(n)+v(n)$, 
where $x(n)$ is the observed signal, $s(n)$ the speech signal, and $v(n)$ the noise signal.
$x(n)$ can be divided into overlapping frames of length $L$, represented by $\mathbf{x}_t\in\mathbb{R}^{L}$,
where $t = 1,\dots,T$ represents the frame index and $T$ represents the total number of frames.
A matrix $\mathbf{X}\in \mathbb{R}^{L\times T}$ can then be formed by concatenating $\mathbf{x}_t$ for all frames $t$.

\subsection{Conv-TasNet}
We will now review Conv-TasNet~\cite{luo2019conv}.
$\mathbf{X}$ is transformed into $N$-dimensional representations  $\mathbf{W}\in\mathbb{R}^{N\times T}$ for all frames by multiplying by a trainable linear encoder $\mathbf{U}\in \mathbb{R}^{N\times L}$ as follows:
\begin{equation}
\mathbf{W} = \mathbf{U}\mathbf{X}.
\label{eq:encoder}
\end{equation}
$\mathbf{W}$ is fed into a TCN-based separation block $\mathcal{F}$, which is composed of a bottleneck layer, TCN blocks, and a mask estimation block~\cite{luo2019conv}, 
\begin{equation}
\mathbf{M}_k = \mathcal{F}_k(\mathbf{W}),
\label{eq:tcn}
\end{equation}
where $\mathbf{M}_k\in \mathbb{R}^{N\times T} (k=1,\dots,K)$
are the masks for $K$ sources.
Then, $\mathbf{M}_k\in \mathbb{R}^{N\times T}$
is multiplied by $\mathbf{W}$,
\begin{equation}
\mathbf{Z}_k = \mathbf{M}_k\odot\mathbf{W},
\label{eq:masking}
\end{equation}
where $\mathbf{Z}_k\in \mathbb{R}^{N\times T}$ is the $N$-dimensional representation of each source and $\odot$ is the Hadamard product.
$\mathbf{Z}_k$ is multiplied by a trainable linear decoder $\mathbf{V}\in \mathbb{R}^{L\times N}$,
\begin{equation}
\hat{\mathbf{S}}_k = \mathbf{V}\mathbf{Z}_k,
\label{eq:decorder}
\end{equation}
where $\hat{\mathbf{S}}_k\in \mathbb{R}^{L\times T}$
contains the estimated frames for each source.
Each estimated source $\hat{s}_k(n)$ is finally reconstructed by overlapping and adding the $T$ columns in $\hat{\mathbf{S}}_k$.
The parameters of Conv-TasNet are learned by minimizing the SI-SNR loss:
\begin{equation}
L_{\text{SI-SNR}} = -\frac{1}{K}\sum_{k=1}^K 10\log_{10}(\|\alpha\mathbf{s}_k\|^2/\|\alpha\mathbf{s}_k-\hat{\mathbf{s}}_k\|^2),
\label{eq:sisnrloss}
\end{equation}
where  $\mathbf{s}_k$ and $\hat{\mathbf{s}}_k$
are vector representations of $s_k(n)$ and $\hat{s}_k(n)$, respectively, and $\alpha = \langle \mathbf{s}_k,\hat{\mathbf{s}}_k \rangle/\|\mathbf{s}_k\|^2$.
Permutation-invariant training (PIT) is utilized to determine an appropriate permutation in terms of the order of sources~\cite{kolbaek2017multitalker}.

There are two options when applying Conv-TasNet to speech enhancement.
The first option is to estimate only the speech signal, corresponding to $K=1$, which obviously does not require PIT.
The second option is to estimate both speech and noise signals, corresponding to $K=2$, defining $k=1$ as speech signal and $k=2$ as noise signal. 
This option also does not require PIT.
In \cite{kinoshita2020improving}, it was shown that the second option gives better speech recognition performance than the first option.
Therefore, we evaluate these options in terms of metrics highly correlated with subjective quality.

\subsection{PASE}
\label{ss:pase}
Although self-supervised learning applied to speech related tasks have gained popularity, there remains a challenge in designing proxy self-supervised learning tasks.  In~\cite{Pascual2019,ravanelli2020multi}, the authors designed a self-supervised learning framework to train a problem-agnostic speech encoder (PASE), a deep neural network based encoder that maps a raw waveform into an encoded speech representation.  The encoded speech representation is learned by having many small multi-layer perceptrons (workers), each taking the encoded speech representation as input and trying to predict a known audio transformation of the original input waveform.  For example, one worker tries to predict the log-power spectra (LPS), another worker tries to predict the Mel Frequency Cepstral Coefficients (MFCCs) and another worker tries to predict the original waveform from the encoded speech representation obtained by PASE.  Since the ground-truth LPS, MFCCs, and waveform can be computed from the original raw waveform, each worker is trained using the mean-squared error between the prediction and the computed ground-truth features.  Both PASE and all the workers are optimized together using backpropagation.  
In \cite{ravanelli2020multi}, they introduced PASE+ where the total number of workers increased to 12. 
We will show in Sec.~\ref{ss:loss} how PASE+ can be used to define a new loss function.

\subsection{Evaluation metrics}
For the evaluation of our systems, we use composite objective measures (i.e., CSIG, CBAK, and COVL)~\cite{hu2007evaluation} and 
the perceptual evaluation of the speech quality (PESQ) measure~\cite{rix2001perceptual}, as they are considered to be highly correlated with subjective quality~\cite{hu2007evaluation}.
We will refer to these metrics as perceptual quality metrics hereafter.
In addition to these metrics, the scale-invariant signal-to-noise ratio (SI-SNR) and segmental SNR (SSNR) are used for multiaspect comparison in some of our experiments.
Note that SSNR is considered to have a lower correlation with subjective quality~\cite{hu2007evaluation}.

\end{spacing}
\begin{spacing}{0.92}
\vspace{-1mm}
\section{Proposed method}

\subsection{STFT-based approach using TCN}
We propose to replace the trainable encoder and decoder $\mathbf{U},\mathbf{V}$ of Conv-TasNet with discrete Fourier basis functions (nontrainable), which makes $\mathbf{W}$ regular complex spectrograms.
Eqs.~\eqref{eq:encoder}-\eqref{eq:decorder} are rewritten as
\begin{equation}
\mathbf{W}_\text{SPEC} = \mathbf{U}_\text{STFT}\mathbf{X},
\quad
\hat{\mathbf{S}}_k = \mathbf{V}_{\text{ISTFT}} \left[\mathcal{F}_k(\mathbf{W}_\text{INPUT})\odot\mathbf{W}_\text{SPEC}\right],
\label{eq:proposed_method}
\end{equation}
where $\mathbf{U}_\text{STFT}$, $\mathbf{W}_\text{SPEC}$, and $\mathbf{V}_{\text{ISTFT}}$ are real-valued matrices, although their first half ($1,\dots,N/2$ in the first dimension) represents real values and their latter half ($N/2+1,\dots,N$ in the first dimension) represents imaginary values of the original complex representation.
We propose two ways of defining $\mathbf{W}_\text{INPUT}$.
The first way is to simply assign $\mathbf{W}_\text{SPEC}$ as $\mathbf{W}_\text{INPUT} \coloneqq \mathbf{W}_\text{SPEC}$.
The second way defines $\mathbf{W}_\text{INPUT} \coloneqq \mathbf{W}_\text{AP}$, where $\mathbf{W}_\text{AP}$ is the real-valued matrix obtained by transforming $\mathbf{W}_\text{SPEC}$ to an amplitude and phase representation.
We also remove the sigmoid function of the mask estimation block of $\mathcal{F}$ such that the masks can take negative values. 
This is important as the masks are applied to the real and imaginary part of the spectrum.
We will refer to this method as STFT-TCN.

\subsection{Low-latency algorithm}
\begin{figure}
  \centering
  \includegraphics[width=7.0cm]{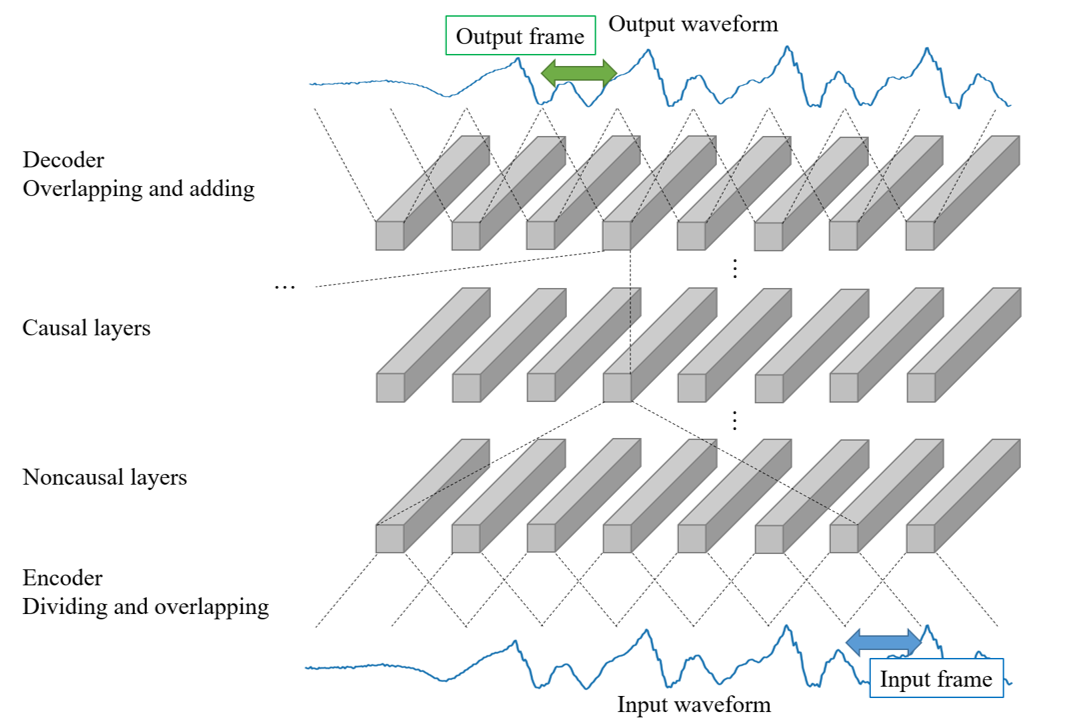}
  \caption{Noncausal layers and causal layers. When the overlapping utilized in the encoder and the decoder is 1/2 and the delay of the noncausal layers is three frames, for instance, the algorithm accesses five future frames.}
\label{fig:nc_layers}
\end{figure}

For the methods based on a TCN, it is reported that 
the network can be easily modified to a low-latency algorithm by 
applying causal convolutions~\cite{bai2018empirical, luo2019conv}.
When access to a limited number of future frames is allowed, as was the case for the DNS Challenge,
it is better to utilize as much future information as possible.
Therefore, we propose to let the first several layers be noncausal and the other layers be causal as shown in Fig.~\ref{fig:nc_layers}.
When the overlapping utilized in the encoder and the decoder is 1/2 and the delay of noncausal layers is three frames, for instance, the output frame at $t = t_{\text{input}}-4$ is computed when the input frame at $t = t_{\text{input}}$ is available, indicating that the algorithm accesses five future frames.

\subsection{Loss functions}
\label{ss:loss}
In addition to the SI-SNR loss from~\eqref{eq:sisnrloss}, 
we use two more loss functions from literature and propose our own loss function based on PASE.

First, in ~\cite{kinoshita2020improving}, it was reported that the classical SNR is also a suitable loss function, which preserves the scale of estimated signal and is calculated as
\begin{equation}
L_{\text{SNR}} = -\frac{1}{K}\sum_{k=1}^K 10\log_{10}(\|\mathbf{s}_k\|^2/\|\mathbf{s}_k-\hat{\mathbf{s}}_k\|^2).
\label{eq:snrloss}
\end{equation}

Second, power-compressed MSE (PCMSE) was utilized in \cite{wisdom2019differentiable,yin2019phasen,erdogan2018investigations}.
To satisfy STFT consistency~\cite{wisdom2019differentiable}, 
we apply the STFT to estimated signals $\hat{\mathbf{S}}_k$
and clean signals $\mathbf{S}_k$, which are created by
dividing, overlapping and concatenating $\hat{\mathbf{s}}_k$ and  $\mathbf{s}_k$ ,
as $\hat{\mathbf{W}}_{\text{SPEC},k} = \mathbf{U}_\text{STFT}\hat{\mathbf{S}}_k, \mathbf{W}_{\text{SPEC},k} = \mathbf{U}_\text{STFT}\mathbf{S}_k$.
Then, the power-compressed MSE loss is calculated as
\begin{equation}
\begin{split}
L_{\text{PCMSE}} = \frac{1}{K}\sum_{f,t}\left[\beta
\left(\left|\hat{W}_k(t,f)\right|^{0.3}-\left|W_k(t,f)\right|^{0.3} \right)^2 \right.\\
\left.+(1-\beta)\left|\hat{W}_k(t,f)^{0.3} - W_k(t,f)^{0.3}\right|^2\right],
\end{split}
\label{eq:compressed_loss}
\end{equation}
where $\hat{W}_k(t,f)$ and $W_k(t,f)$ are complex representations of 
$\hat{\mathbf{W}}_{\text{SPEC},k}$ and $\mathbf{W}_{\text{SPEC},k}$ at frame index $t$ and frequency bin $f$, respectively,
and $W^{0.3}\coloneqq|W|^{0.3}e^{j\angle W}$.

Third, we propose to utilize PASE+~\cite{ravanelli2020multi} as a feature extractor and will refer to the encoded speech representations as PASE features.  The PASE features are calculated as
\begin{equation}
\mathbf{P}_k = \text{PASE}(\mathbf{s}_k),~~
\hat{\mathbf{P}}_k = \text{PASE}(\mathbf{\hat{s}}_k),
\label{eq:pase}
\end{equation}
where $\mathbf{P}_k\in \mathbb{R}^{Q\times R}$ and
$\hat{\mathbf{P}}_k\in \mathbb{R}^{Q\times R}$
are PASE features calculated from clean and estimated signals, respectively. The PASE encoder is mainly composed of several convolutional layers and a quasi-recurrent neural network,
and pretrained by self-supervised learning using clean speech signals as we mentioned in Sec.~\ref{ss:pase}. 
The optimization based on only PASE features cannot estimate the phase of signals accurately as  most of the worker tasks utilized to learn PASE features are independent of phase.
Therefore, we propose to utilize the power-compressed MSE loss simultaneously as
\begin{equation}
L_{\text{PASEMSE}} = \gamma L_{\text{PASE}} + L_{\text{PCMSE}},
\label{eq:pasemse_loss}
\end{equation}
where $L_{\text{PASE}}$ is the mean squared error (MSE) between $\mathbf{P}_k$ and $\hat{\mathbf{P}}_k$.
We will refer to this loss function as PASE-feature MSE (PASEMSE).

\end{spacing}
\begin{spacing}{0.915}
\vspace{-2mm}
\section{Experiment}
We conduct experiments on two different datasets:
the VoiceBank + DEMAND (VBD) dataset\footnote{
\url{https://datashare.is.ed.ac.uk/handle/10283/1942}
}~\cite{veaux2013voice,thiemann2013diverse,valentini2016investigating}, which is relatively small, and the DNS Challenge dataset~\cite{reddy2020interspeech}, which is a large-scale dataset provided by the DNS Challenge.
Several ablation studies are conducted on the VBD dataset, 
and the most promising methods are applied to the DNS dataset.

\begin{figure}[t]
  \centering
  \includegraphics[width=8.0cm]{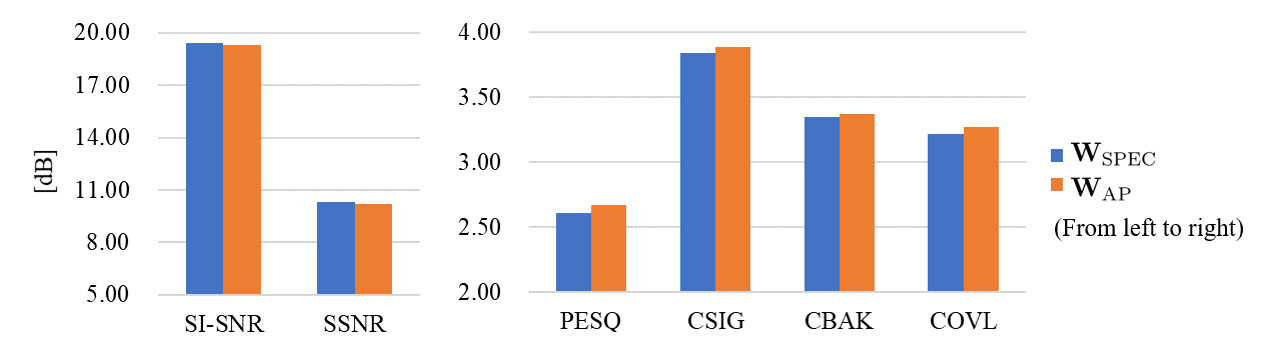}
  \caption{Comparison of the input type of the TCN-based separation block in STFT-TCN.}
\label{fig:ri_ap}
\end{figure}

\begin{figure}[tb]
  \centering
  \includegraphics[width=8.0cm]{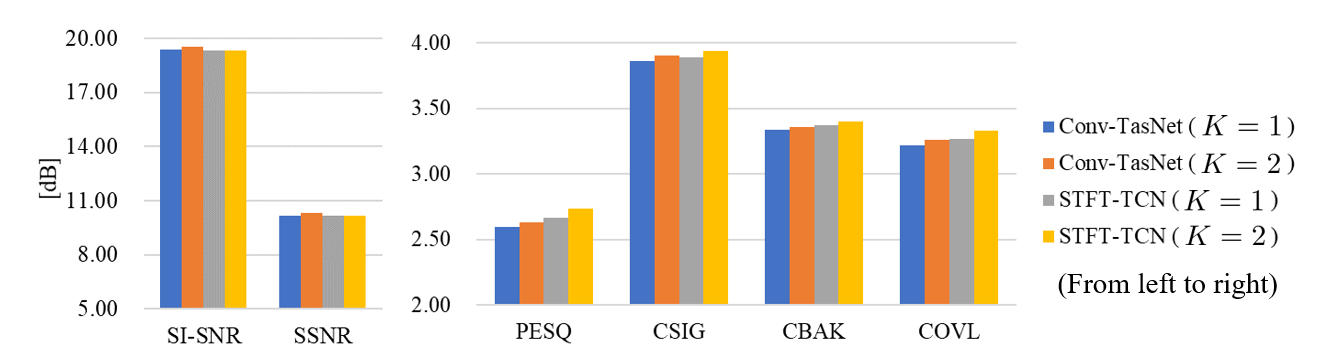}
  \caption{Comparison of the number of output sources $K$.} 
\label{fig:out_comp}
\end{figure}

\begin{table}[tb]
\begin{center}
    \caption{Comparison with existing methods on VBD dataset.}
  \label{tab:all_comp}
\scalebox{0.68}{

    \centering
   \begin{tabular}{l|c|cccc}
    \toprule
        Method & Causality & PESQ & CSIG & CBAK & COVL \\ 
        \midrule
        Noisy & - & 1.97  & 3.35  & 2.44  & 2.63  \\ 
        \midrule
               Open-Unmix~\cite{stoter2019open,uhlich_stefan_2020_3786908} & noncausal & 2.39 & 3.12 & 3.19 & 2.73  \\ 
        DFL~\cite{germain2018speech} & noncausal &  & 3.86  & 3.33  & 3.22  \\ 
        MDPhD~\cite{kim2018multi} & noncausal & 2.70  & 3.85  & 3.39  & 3.27  \\ 
        PHASEN~\cite{yin2019phasen} & noncausal & 2.99  & 4.21  & 3.55  & 3.62  \\ 
        SDR-PESQ~\cite{kim2019end} & noncausal & 3.01  & 4.09  & 3.54  & 3.55  \\ 
        \midrule
        PHASEN (our impl.) & noncausal & 2.58  & 3.91  & 3.20  & 3.23  \\ 
        Conv-TasNet & noncausal & 2.66 & 4.06 & 3.28 & 3.35   \\ 
        Conv-TasNet & 33~ms look ahead & 2.63 & 4.02 & 3.19 & 3.32  \\ 
        Conv-TasNet & 1~ms look ahead  & 2.53 & 3.95 & 3.08 & 3.23  \\ 
        STFT-TCN & noncausal & 2.89 & 4.24 & 3.40 & 3.56  \\ 
        STFT-TCN & 40~ms look ahead & 2.80 & 4.17 & 3.30 & 3.49  \\ 
        STFT-TCN & 4~ms look ahead & 2.73 & 4.11 & 3.25 & 3.42  \\ 
        \bottomrule
    \end{tabular}
}
\end{center}
\end{table}

\begin{figure*}[t]
  \centering
  \includegraphics[width=12.0cm]{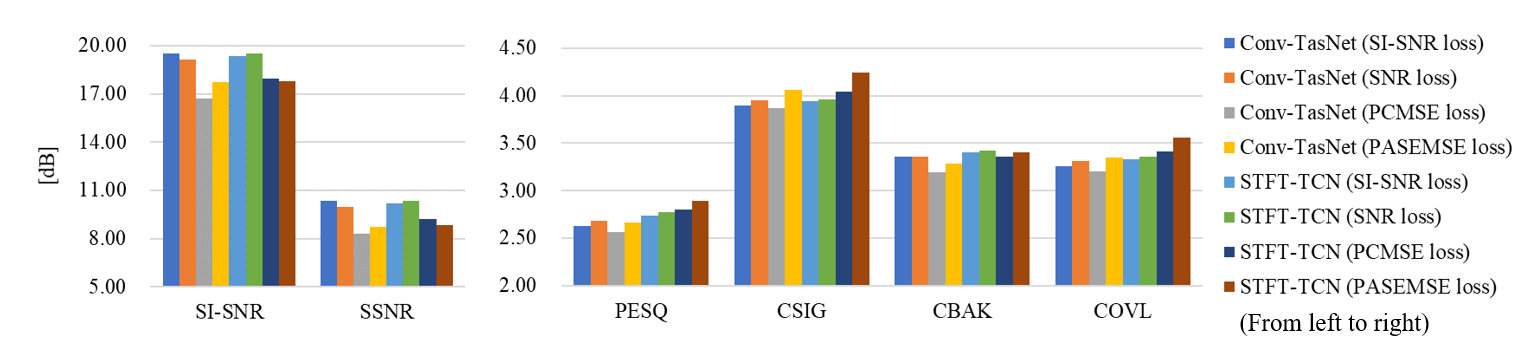}
  \caption{Comparison of loss functions on VBD dataset. The performance of the method using PASEMSE loss is very high in terms of the metrics correlated with subjective quality (PESQ, CSIG, CBAK, and COVL).}
\label{fig:loss_comp}
\end{figure*}

\begin{table*}[htb]
\begin{center}
    \caption{Evaluation results on DNS dataset. Conv-TasNet using SNR loss achieved the best performance.}
      \label{tab:dns_comp}
\scalebox{0.58}{

    \centering
    \begin{tabular}{l|C{1.7cm}|C{1.7cm}|C{1.8cm}|C{1.5cm}|C{1.8cm}|cccc|cccc}
\toprule
    \multirow{2}{*}{Method} & \multirow{2}{1.7cm}{ \# of params [million]} & \multirow{2}{1.7cm}{Time per frame [ms]} & \multirow{2}{1.8cm}{Time to infer a frame [ms]} & \multirow{2}{1.5cm}{Loss function} & \multirow{2}{1.8cm}{Reverberant augmentation} & \multicolumn{4}{c|}{No reverb} & \multicolumn{4}{c}{With reverb} \\ 
        & & & & & &  PESQ & CSIG & CBAK & COVL & PESQ & CSIG & CBAK & COVL \\ 
        \midrule
        Noisy (Raw) & - & - & - & - & - & 1.58  & 3.03  & 2.53  & 2.27  & 1.82  & 3.50  & 2.80  & 2.64  \\
        \midrule
        Baseline (NSNet)~\cite{xia2020weighted} & 1.26 & 10.0 & - & \cite{xia2020weighted} & \cite{xia2020weighted} & 1.83  & 2.88  & 2.75  & 2.32  & 1.52  & 2.71  & 2.51  & 2.09  \\
        \midrule
        STFT-TCN & 5.03 & 4.0 & 350.1 & PASEMSE & \checkmark &1.51 & 2.99 & 2.25 & 2.22  &1.51 & 3.20 & 2.31 & 2.33  \\ 
        STFT-TCN & 5.03 & 4.0 & 348.1 & PCMSE &  & 2.24  & 3.63  & 3.12  & 2.93  & 1.46  & 2.89  & 2.45  & 2.15  \\ 
        STFT-TCN & 5.03 & 4.0 & 353.8 & PCMSE & \checkmark & 2.16  & 3.54  & 3.04  & 2.84  & 2.19  & 3.75  & 3.10  & 2.97  \\ 
        Conv-TasNet & 5.08 & 1.0 & 355.5 & PASEMSE & \checkmark & 1.97 & 3.49 & 2.88 & 2.71 & 2.11 & 3.77 & 3.02 & 2.92   \\         
        Conv-TasNet & 5.08 & 1.0 & 355.6 & SNR & & 2.72  & 4.05  & \bf{3.65}  & 3.39  & 1.85  & 3.25  & 2.78  & 2.53  \\ 
        Conv-TasNet (*) & 5.08 & 1.0 & 351.9 & SNR & \checkmark & \bf{2.73}  & \bf{4.07}  & 3.64  & \bf{3.41}  & \bf{2.71}  & \bf{4.21}  & \bf{3.67}  & \bf{3.47}  \\ 

        \bottomrule
    \end{tabular}
    }
\end{center}
\vspace{-6mm}
\end{table*}

\subsection{VoiceBank + DEMAND dataset}
In the VBD dataset, 11,572 and 824 noisy-clean speech pairs are provided as the training and test set, respectively.
We extract 300 pairs from the training set and utilize them as the validation set.
In Conv-TasNet, $L=32$ and $N=512$ are applied with 1/2 overlapping, and other hyperparameters are those of the best configuration in the original paper~\cite{luo2019conv}.
In STFT-TCN, $L=192$ and $N=512$ are applied with 2/3 overlapping such that the future information that the algorithm accesses is just 40~ms when the number of noncausal layers is three.
To calculate the loss function \eqref{eq:compressed_loss} and \eqref{eq:pasemse_loss}, $\beta=0.5$ and $\gamma=0.25$ are used, which are the best configuration in our preliminary experiment.
All networks in our experiment are trained with the Adam optimizer~\cite{kingma2014adam}.
The learning rate is initially set to 0.001 and halved if the output of the loss function in the validation set is not improved in three consecutive epochs.

We first compare two types of $\mathbf{W}_\text{INPUT}$ in \eqref{eq:proposed_method}, namely, $\mathbf{W}_\text{SPEC}$ and $\mathbf{W}_\text{AP}$, in terms of STFT-TCN (SI-SNR is assigned for the loss function and $K=1$).
As shown in Fig.~\ref{fig:ri_ap}, 
the performance of the method using $\mathbf{W}_\text{AP}$ is slightly better than that of the method using $\mathbf{W}_\text{SPEC}$ in terms of perceptual quality metrics (i.e., PESQ, CSIG, CBAK, and COVL).
On the basis of this result, hereafter $\mathbf{W}_\text{AP}$ is utilized as $\mathbf{W}_\text{INPUT}$ in STFT-TCN.

Next, the number of output sources, $K$, which determines whether the noise signal is estimated simultaneously or not, is compared for both Conv-TasNet and STFT-TCN.
Both are trained with the SI-SNR loss \eqref{eq:sisnrloss}.
Fig.~\ref{fig:out_comp} shows that 
the performance of both methods with $K=2$ is consistently higher than that with $K=1$ in terms of all metrics as was also noted in \cite{kinoshita2020improving}.
Thus, learning the noise mask together with the speech mask improves performance and we adopt $K=2$ hereafter.
It is also shown that
STFT-TCN surpasses Conv-TasNet in terms of perceptual quality metrics.

Fig.~\ref{fig:loss_comp} shows the result of comparing loss functions for Conv-TasNet and STFT-TCN.
It also shows that SNR loss is more suitable than SI-SNR loss for the perceptual quality metrics, which is a similar tendency to that with \cite{kinoshita2020improving}.
PCMSE can improve the performance of STFT-TCN in terms of metrics correlated with subjective quality, while it does not work well for Conv-TasNet.
The most important finding in Fig.~\ref{fig:loss_comp} is that 
the performance of the method using PASEMSE is very high for the perceptual quality metrics.

Finally, we compared our proposed methods using PASEMSE with other existing methods whose performance on the VBD dataset was previously reported.
As shown in Table~\ref{tab:all_comp},
STFT-TCN outperforms Conv-TasNet, DFL~\cite{germain2018speech}, and MDPhD~\cite{kim2018multi} in terms of all perceptual quality metrics.
From the table, it can be seen that STFT-TCN performs slightly worse than PHASEN~\cite{yin2019phasen}. We implemented PHASEN ourselves -- see ``PHASEN (our impl.)'' in Table~\ref{tab:all_comp} -- but could not reproduce the reported results. It seems that PHASEN is sensitive to the chosen hyperparameter values.
In addition, STFT-TCN surpassed SDR-PESQ~\cite{kim2019end}, which uses PESQ for its loss function, in terms of CSIG and COVL.
Note that our proposed methods achieve such performance without training explicitly for PESQ. 
Moreover, it was proven that future information obtained by the noncausal layer contributes to the performance even if the range is very short such as 33~ms (Conv-TasNet with five noncausal layers) or 40~ms (STFT-TCN with three noncausal layers), and they are still comparable to other state-of-the-art methods.

\subsection{DNS Challenge dataset}
In the DNS dataset, over 60,000 speech and noise samples are provided as the training set.
Synthetic and real recorded test sets are also provided, and the synthetic set is further divided into ``No reverb'' and ``With reverb'' sets, according to whether reverberation exists.
We focus on the synthetic set for the quantitative evaluation.
We first roughly screened out noisy speech using a speech enhancement model trained in our preliminary experiment.
Then, 150 speech and noise data are extracted from the training set and utilized for the validation set.
To adapt the model to the ``With reverb'' set, we augment the speech data by convolving the impulse response generated by the image-source method~\cite{lehmann2008prediction}.
The reverberation time T60 is randomly sampled from 0.2~s to 0.8~s.
Augmented speech and noise are randomly and independently sampled and mixed to create pairs of data for the training (i.e., on the fly~\cite{erdogan2018investigations}), and 10,000 iterations are defined as 1 epoch.
As the algorithm is allowed to access only 40ms of future frames in the DNS Challenge, we evaluate Conv-TasNet with access to 33~ms of future frames and STFT-TCN with access to 40~ms of future frames introduced in Table~\ref{tab:all_comp}.
All other configurations are the same as those in the experiment on the VBD dataset.

Table~\ref{tab:dns_comp} shows the evaluation results on DNS dataset.
We implement our methods with PyTorch~\cite{paszke2017automatic} as each frame is processed\footnote{
\url{https://github.com/ykoyama58/tcnse}} (i.e., ``1-frame-input-1-frame-output'')
 and measure the time to infer a frame using an Intel Core i5-6200U clocked at 2.4 GHz by taking the average over the whole utterance of the first sample in the test set. 
We find that the data augmentation using impulse responses significantly improves the performance on the ``With reverb'' set, while a slight degradation in STFT-TCN can be found on the ``No reverb'' set.
The most notable finding is that the method based on the Conv-TasNet outperforms STFT-TCN on the DNS dataset but not on the VBD dataset.
This can be interpreted to mean that the large training dataset enables the network to learn appropriate encoders and decoders.
Also, the method using PASE does not work well on the DNS dataset.
We conclude that STFT-TCN and the loss function using PASE are effective only for smaller dataset.

Since the computational complexity of our method does not satisfy the requirement of the real-time track, we have submitted our best method (marked in Table~\ref{tab:dns_comp} by *) to the non-real-time track in the DNS Challenge.

\section{Conclusions}
We proposed and evaluated two TCN-based approaches, Conv-TasNet and STFT-TCN, on two different datasets, VBD and DNS datasets.
For the VBD dataset, STFT-TCN utilizing PASE for the loss function outperformed Conv-TasNet and other existing approaches in terms of perceptual quality metrics, while Conv-TasNet surpassed STFT-TCN on the DNS dataset.
In future work, we will explore how to shrink our networks and implement them in a real-time system.

\end{spacing}

\bibliographystyle{IEEEtran}

\bibliography{mybib}

\end{document}